\documentclass[prd,aps,preprintnumbers,twocolumn,floatfix,nofootinbib]{revtex4-2}
\usepackage{fix-revtex-4-2-bib}
\usepackage{qcd}
\usepackage{epsfig}
\usepackage{amsmath}
\usepackage{graphicx,xcolor}
\usepackage[normalem]{ulem}
\usepackage{hyperref}

\begin{document}

\preprint{JLAB-THY-23-3909}

\title{On the definition of fragmentation functions and the violation of sum rules}

\author{John Collins}
\email{jcc8@psu.edu}
\affiliation{Department of Physics, Penn State University, University Park PA 16802, USA}
\affiliation{\\ \href{https://orcid.org/0009-0000-8494-6992}{ORCID: 0009-0000-8494-6992}}
\author{Ted~C.~Rogers}
\email{trogers@odu.edu}
\affiliation{Department of Physics, Old Dominion University, Norfolk, VA 23529, USA}
\affiliation{Jefferson Lab, 12000 Jefferson Avenue, Newport News, VA 23606, USA}
\affiliation{\\ \href{https://orcid.org/0000-0002-0762-0275}{ORCID: 0000-0002-0762-0275}}
\date{January 14, 2024}

\begin{abstract}
We point out a problem with the formulation and derivations of sum rules for quark
fragmentation functions that impacts their validity in QCD, but which potentially points toward an improved understanding of final states in inclusive hard processes. Fragmentation functions give the distribution of final-state hadrons arising from a parton exiting a hard scattering, and the sum rules for momentum, electric charge, etc express conservation  of these quantities.  The problem arises from a mismatch between the quark quantum numbers of the initial quark and the fact that all observed final-state hadrons are confined bound states with color zero. We point that, in a confining theory like QCD, the Wilson line in the operator definition of a fragmentation function entails that the final state in a fragmentation function includes a bound state in the external field generated by the Wilson line.  We justify this with the aid of general features of string hadronization.  The anomalous bound states are restricted to fractional momentum $z=0$.  They tend to invalidate sum rules like the one for charge conservation when applied to the fragmentation functions inferred from experimental data, but not the momentum sum rule. We propose to exploit our ideas in future studies as a way to relate the ffs extracted from inclusive cross sections to more detailed non-perturbative descriptions of final state hadronization. We also describe scenarios wherein the traditional sum rules might remain approximately valid with a reasonably high degree of accuracy.  
\end{abstract}

\maketitle

\enlargethispage{10mm} 

\section{Introduction}
Factorization theorems for inclusive processes in QCD are the theoretical basis for much  existing QCD phenomenology.   When factorization is applied to distributions of final-state hadrons, fragmentation functions (ffs) are used for the distributions of hadrons that arise from  partons exiting the hard scattering.  In analogy with sum rules for parton densities, sum rules for ffs have been written down~\cite{Collins:1981uw,Vendramin:1981te,Majumder:2004br,deFlorian:2003cg,Meissner:2010cc,Pitonyak:2023gjx}, including for cases with more than one detected final-state hadron, and with proofs that ostensibly apply non-perturbatively. 
The momentum sum rule in particular is sometimes proposed as a constraint in phenomenological extractions~\cite{Hirai:2007cx,deFlorian:2014rmz}.
Ideally, one could combine the use of ffs in studies of scattering data with theoretical non-perturbative studies of hadronization. However, one finds that a contradiction arises concerning the final state sum/integral that appears in the operator ff definition, \eref{basicdef2} below. 

If we ignore complications, the definition of an ff involves an initial partonic state created by applying a light-front creation operator for a quark (or gluon) 
field to the vacuum.  Then an ff is a distribution of 
particles in the final state (i.e., at asymptotically large time). The sum rules are just expressions of conservation laws in QCD. 
Now, all observed final-state particles in QCD have hadronic quantum 
numbers.  In contrast, the initial state of a quark ff has 
quark quantum numbers, e.g., fractional baryonic and electric charge, and so 
the final state must also have the same quantum numbers.  This not possible for a normal hadronic final state. 

In this paper, we propose a resolution of this 
paradox.  It involves the Wilson line needed for a gauge-invariant 
definition of the operator creating the partonic state.  Since the Wilson line 
extends out to infinity, it requires a modified concept for parts of 
the final state.  By using the successful string scheme for non-perturbative 
hadronization, we find that the final state of a quark ff should contain a particle that is effectively the bound state of quark(s) with the Wilson line, but that this is localized essentially at zero fractional momentum, $z=0$. This does not affect factorization 
and does not contribute to the momentum sum rule.  But it modifies how other sum rules are to be applied phenomenologically.

Factorization applies in a general quantum field theory, not just QCD.  So part of our paper will contrast the situation in a QCD-like theory with the situation in a non-gauge model theory where no Wilson lines are used and there is no quark confinement.

\section{The definition} 
The standard definition of a ``bare'' quark ff $d_{(0),h/j}(z,\T{p}{})$ is (e.g., Eq.~(12.35) of \cite{Collins:2011qcdbook}):
\begin{multline}
d_{(0),h/j}(z,\T{p}{}) ~ \langle j, k_1 |j, k_2 \rangle \\
\equiv \frac{\sum_X \langle j, k_1 | h, X, \text{out} \rangle \langle h, X, \text{out} |j, k_2 \rangle}{2 z (2 \pi)^{3-2 \epsilon}}  \, .
\label{e.basicdef}
\end{multline}
Here, $| j, k_1 \rangle$ and $|j, k_2 \rangle$ are initial quark states with flavor $j$, which are obtained by applying a light-front creation operator to the vacuum. We define light-front variables by $k^{\pm}= (k^0 \pm k^z) / \sqrt2$, and a two dimensional transverse momentum.  The initial quark states have given values of $k_1^-$ and $k_2^-$, and transverse momenta (with the transverse momenta taken to zero later).  The variable $z$ is the fractional minus component of the quark momentum carried by the observed final state hadron $h$, $p^- = z k^-$, and $\T{p}{}$ is its transverse momentum.  The out-states are $|h,X, \text{out} \rangle$ with $h$ labeling the measured particle species and the sum/integral over $X$ representing a complete sum over all other contributions to the final state.  Because states of definite momentum are non-normalizable, we work indirectly to get a number density, which entails the factor $\langle j, k_1 |j, k_2 \rangle$ on the left of \eref{basicdef}.

We start by considering nongauge theories, so we have omitted from \eref{basicdef} the color factor in the corresponding equation for QCD given in \cite{Collins:2011qcdbook}. We require that the theory be renormalizable and, to show the contrast with QCD, we require it to contain elementary Dirac fields that we call quarks, with at least one flavor.  
The definition is stated in $4 - 2 \epsilon$ dimensions to anticipate the existence of ultraviolet divergences that need to be regulated and renormalized. Equation \eqref{e.basicdef} actually defines a bare transverse-momentum-dependent ff. The bare \emph{collinear} ff is obtained by integrating over all $\T{p}{}$. It can be re-expressed in the more familiar form,
\begin{align}
&{} d_{(0),h/j}(z) \equiv \int \diff[2 - 2 \epsilon]{\T{p}{}} d_{(0),h/j}(z,\T{p}{}) \no
= &{} \frac{{\rm Tr}_D}{4} 
\sum_X z^{1 - 2 \epsilon} \int \frac{\diff{x^+}}{2 \pi} 
e^{i k^- x^+} \gamma^{-} \times \no 
&{} \, \times \langle 0 | \psi_j^{(0)}\parz{x/2} | h,X, \text{out} \rangle \langle h, X, \text{out} | \overline{\psi}_j^{(0)} \parz{-x/2} | 0 \rangle \, ,
\label{e.basicdef2}
\end{align}
with the fields separated in the $+$-direction.
Some of the manipulations needed to give \eref{basicdef2} are just  to deal with the fact that partonic states with definite $k^-$ and $\T{k}{}$ are not normalizable, so the definition of probability densities in terms of quantum mechanical states is obtained indirectly.
In addition, there is a Lorentz transformation, to take the transverse momentum of the hadron to zero while preserving the minus-components of momenta. 

Finally, we define $|j,k\rangle = b_{k,j}^\dagger |0\rangle$, where $b_{k,j}^\dagger$ is a light-front creation operator.  It is obtained from an expansion of the Fourier transform of the good components of the quark field on a light-front, and obeys anticommutation relations
\begin{equation}
\left\{b_{k,\alpha}, b^\dagger_{l,\alpha'} \right\} = (2 \pi)^3 2 k^- \delta\parz{k^- - l^-} \delta^{(2)}\parz{\T{k}{}- \T{l}{}} \delta_{\alpha \alpha'} \, .
\end{equation}
In terms of the quark field of a specific flavor, the $b_{k\alpha}$ operator, for example, is 
\begin{equation}
\label{e.lf.expand}
b_{k,\alpha}(x^+) = \int \diff{x^+} \diff[2]{\T{x}{}} e^{ik^- x^+ - i \T{k}{} \cdot \T{x}{}} \bar{u}_{k,\alpha} \gamma^- \psi(x) \, .
\end{equation}
See, for example, Sec.\ 6.6 of \cite{Collins:2011qcdbook} for a more detailed review of light-cone quantization. Our reason for discussing it here is to emphasize that there are two different types of particle state involved in ffs: those like the fragmenting quark states $| j, k \rangle$, and those for stable particles in the final states at large time scales, with correspondingly different operators to create them.

\section{Sum rules}
Now we briefly review the essential steps in the derivation of the momentum sum rule for ffs. For a more complete treatment see Sec.\ 5.4 of \cite{Collins:1981uw}.

We start with the following expression for the single inclusive final state sum in terms of creation and annihilation operators for out-particles:~\footnote{We emphasize the need to make clear conceptual and notational distinctions between the (off-shell) light-front creation operators and those for on-shell asymptotic final-state particles, and similarly for the states they create.  The issue is particularly acute in a theory without quark confinement, where one finds a non-trivial ff for a quark to a quark, as we will illustrate calculationally in the Appendix.  Such a concept is paradoxical if the use of two different types of quark state is not made explicit.} 
\begin{align}
 &\sum _X | h,X, \text{out} \rangle \langle h, X, \text{out} | \no
 & \qquad \equiv \sum_X a_{h,p,\text{out}}^\dagger  | X,\text{out} \rangle \langle X,\text{out} | a_{h,p,\text{out}}\no
 & \qquad = a_{h,p,\text{out}}^\dagger a_{h,p,\text{out}} \, ,
\label{e.final_state_def}
\end{align}
so that, for example, \eref{basicdef2} is equivalent to
\begin{multline}
d_{(0),h/j}(z) 
= \frac{{\rm Tr}_D}{4} 
z^{1 - 2 \epsilon} \int \frac{\diff{x^+}}{2 \pi} 
e^{i k^- x^+} \gamma^{-} \times \\
\times \langle 0 | \psi_j^{(0)}\parz{x/2} a_{h,p,\text{out}}^\dagger a_{h,p,\text{out}} \overline{\psi}_j^{(0)} \parz{-x/2} | 0 \rangle .
\end{multline}%

In terms of $a_{h,p,\text{out}}^\dagger$ and $a_{h,p,\text{out}}$, the operators for components of momentum are
\begin{equation}
\mathcal{P}^\mu = 
 \sum_h \int_0^\infty \frac{\diff{p^-}}{2 p^-} 
     \int \frac{\diff[2 - 2 \epsilon]{\T{p}{}}}{(2 \pi)^{3-2\epsilon}} \, a_{h,p,\text{out}}^\dagger p^\mu a_{h,p,\text{out}} \, , \label{e.mom_op}
\end{equation}
where the sum over $h$ is over all species and spin states of stable particle.  We substitute this into the matrix element $\langle j,k_1| \mathcal{P}^- |j,k_2\rangle$, which equals $k_1^-\langle j,k_1|j,k_2\rangle$, and then apply \eref{final_state_def} to get an integral over $z$ and $\T{p}{}$ of the right-hand side of \eref{basicdef}.  This gives the momentum sum rule
\begin{equation}
\label{e.the_sum_rule}
\sum_h \int_0^1 \diff{z} \, z \, d_{(0),h/j}(z) = 1 \, ,
\end{equation}
for each quark flavor $j$.

The remaining nontrivial step is to show that \eref{the_sum_rule} is preserved after renormalization. It is well known that this works for standard renormalization schemes like $\msbar$, so from here forward we will drop ``$(0)$'' subscripts in equations like \eref{the_sum_rule} and assume that renormalization has been implemented. 
 
\section{The final states}
The above sum rule derivation relies, for its validity, on the use of a complete set of basis states $|X,\text{out}\rangle$ whose sum/integral obeys
\begin{equation}
\sum_X |X, \text{out}\rangle \langle X, \text{out}| = \widehat{1} \label{e.unitarity}
\end{equation}
where $\widehat{1}$ is the unit operator on the physical state space $\mathcal{E}$.\footnote{The non-vacuum part of the space is then stratified in a one-particle inclusive form as in \eref{basicdef2}, with $h$ being summed over all possible kinds of single-particle state, including a spin sum.}
The derivation also relies on the sum over $h$ in \eref{basicdef2}, etc, being over all kinds of possible stable single-particle states, such that the momentum operators obey \eref{mom_op}.

Now all observed final-state particles in QCD have integer baryon number and electric charge.  But the initial state $b^\dagger|0\rangle$ in a quark ff has quark quantum numbers, notably fractional electric charge, and so is orthogonal to all purely hadronic final states.  This appears to give zero for the matrix elements in \eref{basicdef2} and hence for the ffs.

The paradox does not arise in a non-gauge theory: There, we can apply a locally smeared quark field to the vacuum to create a normal physical state with quark quantum numbers.  By general principles of QFT, there must be final states of quark quantum numbers, and hence at least one stable particle (bound or not) of the appropriate quantum numbers, including electric charge. 

But in a gauge theory, local quark fields are not gauge-invariant physical operators. Applied to the vacuum, they do not give an unambiguously physical state.  Instead, the field used to define the ff of a quark is multiplied by a Wilson line, which goes out to infinity in an appropriate light-like or almost light-like direction.\footnote{Note that when one treats transverse-momentum dependent ffs there are some complications associated with the details of the Wilson line --- see Ch.~13 of \cite{Collins:2011qcdbook} and references therein. These complications do not affect the basic ideas being explained here.} A Wilson line is effectively a source of color charge, so a Wilson line going out to infinity changes the nature of the possible final-state particles. In a confining theory like QCD, we must have, in addition to normal hadrons, states that are bound to the Wilson line.  For a quark ff, the Wilson line is in a color anti-triplet representation.  So we can have a meson-like state with a quark bound to the Wilson line in a color-singlet configuration. Other possibilities include an antibaryon-like state of two antiquarks bound to the Wilson line, again in a color-singlet configuration.  Since the Wilson line has a rapidity with respect to the fragmenting quark that is infinite, or at least large, we can expect the bound states to be at a fractional momentum $z$ that is zero, or close to zero.

We therefore extend the normal QCD state space $\mathcal{E}$, with its Fock basis of out states, to include these extra bound states.  That is, we replace the hadronic state space $\mathcal{E}$ by
\begin{equation}
    \mathcal{E} \otimes \mathcal{B},
\end{equation}
where $\mathcal{B}$ is the space of bound states for the Wilson line.  Correspondingly there is a modified momentum operator, with corresponding consequences for the sum rules for ffs.

Let $H$ be the set of kinds of normal hadrons, and $B$ be the set of bound states of the Wilson line. We propose that the momentum operator should be modified from the one given in \eref{mom_op} to
\begin{equation}
\mathcal{P}^\mu = \mathcal{P}_{H}^\mu + \mathcal{P}_{B}^\mu \, , 
\end{equation}
where
\begin{align}
   \mathcal{P}_{H}^\mu & \equiv 
   \sum_{h \in H}  \int_0^\infty \frac{\diff{p^-}}{2 p^-} 
   \int \frac{ \diff[2 - 2 \epsilon]{\T{p}{}} }{ (2 \pi)^{3-2\epsilon} }          a_{h,p,\text{out}}^\dagger p^\mu a_{h,p,\text{out}} \, ,
\\
   \mathcal{P}_{B}^\mu & \equiv 
   \sum_{b \in B}  \int_0^\infty \frac{\diff{p^-}}{2 p^-} 
   \int \frac{ \diff[2 - 2 \epsilon]{\T{p}{}} }{(2 \pi)^{3-2\epsilon}}a_{b,p,\text{out}}^\dagger p^\mu a_{b,p,\text{out}} \, .
\end{align}

The conflict about the initial state with quark quantum numbers relies on the theory being QCD-like, with color confinement. So we should expect similar issues to arise in QED in low space-time dimensions ($1+1$, $2+1$), where the classical Coulomb potential rises linearly or logarithmically with distance.  In contrast, QED in $3+1$ dimensions does not have electron confinement, so that electrons do appear as possible final-state particles. Even so, the definition of an electron ff in QED still needs a Wilson line going out to infinity, for which bound states can exist.\footnote{The need for a Wilson line is likely to be related to the complications in defining states of charged particles in QED --- for a review, see Ref.~\cite{Duch:2023xyb}.} So some version of the QCD issues does arise there, probably only in a minor way. 

Once we have modified the momentum operator, we get a corresponding modified form for the momentum sum rule, compared with \eref{the_sum_rule}:
\begin{equation}
\label{e.the_sum_rule2}
\sum_{h \in H} \int_0^1 \diff{z} \, z \, d_{h/j}(z)
= 1 
  - \sum_{b \in B} \int_0^1 \diff{z} \, z \, d_{{b}/j}(z) \, .
\end{equation}
The left-hand side is the same quantity as before, and corresponds to the ffs that can be inferred from scattering data, with their purely hadronic final-state particles.  The second term on the right-hand side represents a deficit with respect to the standard value.  To the extent that the parts of the final-state with a particle of quark quantum numbers give a term that is a delta function at $z=0$, which is the natural expectation, the deficit term is zero, because of the explicit factor of $z$ in the integrand.

Observe that in a non-confining theory, the term with final-state particles of quark quantum numbers is not restricted to $z=0$; indeed it can in part give a term proportional to $\delta(z-1)$, as in perturbative calculations in model QFTs.  Then the set $H$ should be defined to include such terms, and the set $B$ is to be restricted to bound states with the Wilson line.

The situation changes for the flavor sum rules, such as were formulated in \cite{Collins:1981uw}.   For the charge sum rule, our modified derivation gives.
\begin{equation}
\label{e.charge_sum_rule}
\sum_{h \in H} \mathcal{Q}_h \int_0^1 \diff{z} \, d_{h/j}(z) 
= \mathcal{Q}_j 
   - \sum_{b \in B} \mathcal{Q}_{b} \int_0^1 \diff{z} \, d_{{b}/j}(z) \, .
\end{equation}
A $\delta(z)$ term for $d_{b/j}(z)$ no longer gives zero, and we can no longer expect the original formulation \cite{Collins:1981uw} of the sum rule to be necessarily accurate.  A closer and non-perturbative analysis of the dynamics is needed to get a prediction for the right-hand side; this we can do with the aid of the string model, to a useful approximation.  

For the total hadron number sum rule~\cite{Majumder:2004br}, we similarly have
\begin{equation}
\label{e.hadron_sum_rule}
\sum_{h \in H} \int_0^1 \diff{z} \, d_{h/j}(z) = \langle N \rangle - \sum_{b
 \in B} \int_0^1 \diff{z} \, d_{{b}/j}(z) \, ,
\end{equation}
where $\langle N \rangle$ is the average multiplicity of \emph{all} final states produced by the fragmenting quark, including those with nonhadronic quantum numbers.\footnote{It is a slight abuse of language to refer to \eref{hadron_sum_rule} as a ``sum rule'' since the  multiplicity associated with a single ff is not a known measurable or conserved quantity. Moreover, its exact numerical value depends on an arbitrary choice of renormalization scheme or of a cut-off. We will continue to call it the ``hadron number sum rule,'' however, to remain consistent with existing literature, where identities like \eref{hadron_sum_rule} guide interpretations.}  

In the Appendix, we illustrate how the sum rules apply in a renormalizable nongauge theory. The examples there indicate the importance of keeping terms with final-state particles that have quark quantum numbers.

\section{Interpretation in QCD} 
The nature of the final state in a quark ff in QCD is illustrated by the strong-model account of hadronization in $e^+e^-$-annihilation.  We present this as a qualitative perturbative schema and approximation in \fref{epem}, as in the paper by Casher, Kogut and Susskind \cite{Casher:1974vf}.  At center-of-mass energy $Q$, an electron and positron annihilate over a short distance scale $1/Q$ to make an outgoing quark-antiquark pair.  A color field between them, that ends up as a flux tube, is created by gluon emission.  Quark-antiquark pairs are generated in the flux tube, and recombine into color-singlet mesons.  In space-time, the meson production is  roughly localized around a space-like hyperbola. The slowest mesons, of low center-of-mass rapidity, are formed first, and the fastest, high rapidity, mesons are formed on a long time-dilated scale. The result is a two jet structure, with each jet created from one or other of the quark and antiquark.  Between the leading particles in each jet, mesons fill in the rapidity region with an approximately uniform distribution in rapidity.

\begin{figure}
  \includegraphics[width=0.4\textwidth]{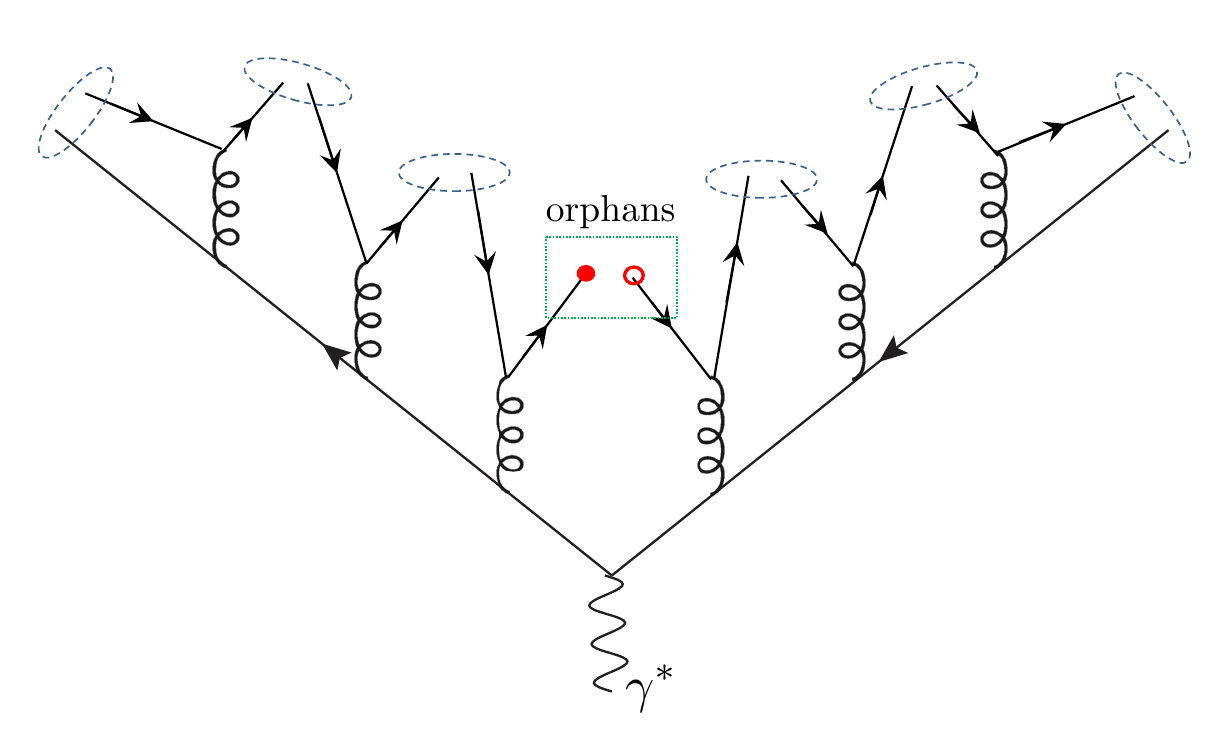} 
\caption{A schematic picture of the production of a final state in the process $e^+ e^- \to \text{hadrons}$. Hadrons on the left hand side move with large negative rapidity, and those on the right hand side move with large positive rapidity. To split the figure into two 
ffs like \fref{ffs_orphan} with quark number 1, the central boxed hadron must be split into an orphan quark and an orphan antiquark.}
\label{f.epem}
\end{figure}

The quark generating a jet preferentially ends up in the leading particle in its jet, with a corresponding bias in the leading particle's charge.  For example, the leading particle in a $u$-quark jet is more often a $\pi^+$ than a $\pi^-$.  But there is, at the same time, no hadron of the fractional charge of the $u$-quark.  This contrasts with the situation in the simple, low-order model in the Appendix, of fragmentation in a non-gauge model.  In that model, the dominant leading particle in the fragmentation of a quark-analog is exactly an on-shell particle of quark quantum numbers. 

The distribution of hadrons in each jet is given by the corresponding fragmentation function.  One can propose splitting the final state in \fref{epem} between the two jets.  For symmetry, this split is in the middle of the box labelled ``orphans'', thereby leading to a left-over quark or antiquark in each part, which we call an orphan quark or antiquark.  This is illustrated in \fref{ffs_orphan}.  In this approximation, whether the orphan is a quark or an antiquark is fully determined by the corresponding property of the jet-initiating parton.  This gives a long-range correlation, given that the orphan is at low momentum.  But no such correlation applies to which flavor the orphan has (e.g., $u$ versus $d$).

\begin{figure}
  \includegraphics[width=0.4\textwidth]{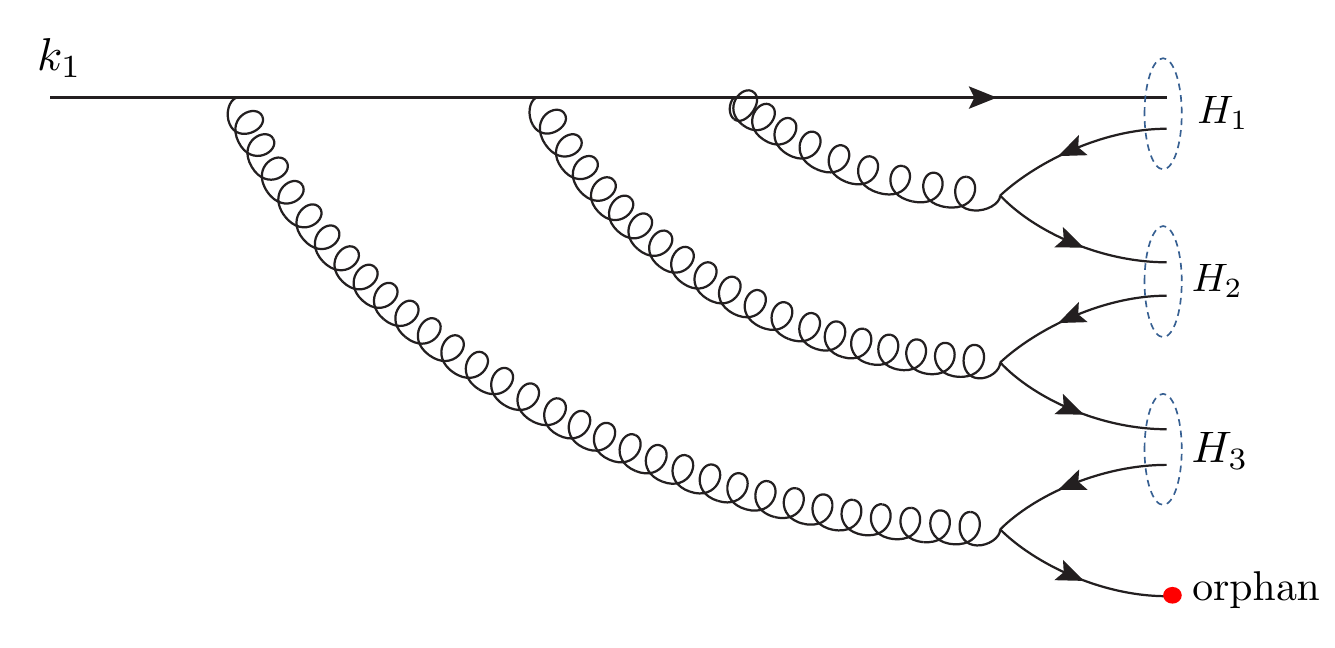} 
\caption{A schematic picture of an allowed final state in a calculation of the quark ffs. Hadrons are labeled by
$H_1$ through $H_3$. These belong in the set of particles labeled $H$ in the text. A final orphan quark with very small $z$, labeled by a red dot, always remains. It is part of the set of states labeled $B$.}
\label{f.ffs_orphan}
\end{figure}

In \fref{ffs_orphan}, only mesons are shown in the final state.   More complex arrangements with multiple leftover quarks and antiquarks can arise to give, for example, baryons in the final state, as in \fref{baryon}.

\begin{figure}
  \includegraphics[width=0.4\textwidth]{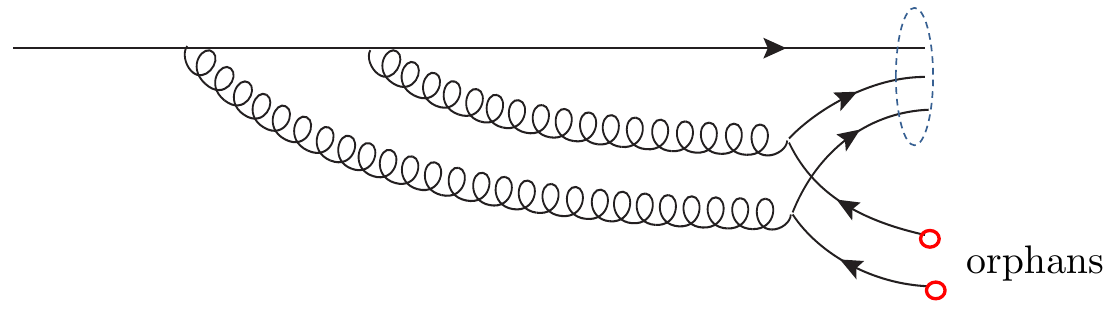} 
\caption{Like \fref{ffs_orphan} but with a baryon in the final state, represented by the oval. The result now has a pair of orphan antiquarks.}
\label{f.baryon}
\end{figure}

Note that the quantitative description of the hadron distribution by a fragmentation function should only be accurate for the faster hadrons, and the just-mentioned split is at the lowest end of the $z$ range for which fragmentation is relevant.  Given a process at some energy, we can choose a $Q$-dependent value $z_{\rm min}$, below which we do not attempt to describe hadron distributions by a fragmentation function of a given parton.   This implies that at a given value of $Q$, the portion of the integrals in the sum rules in the range $0<z<z_{\rm min}$ is not accessible to experimental measurements.  This range decreases approximately like $1/Q$ as $Q$ gets large. The orphan quark or antiquarks at low $z$ carry quantum numbers that are able 
to leak out of the valid factorization region $z > z_{\rm min}$. Thus, we will call the $d_{{b}/j}(z)$ that appear in the correction terms for sum rules like \erefs{charge_sum_rule}{hadron_sum_rule} the ``deficit'' ffs.

We can also apply the string model to the actual ff we defined, that includes a Wilson line.  This is shown in \fref{bound}.  The string model indicates that the orphan quark should simply combine with the Wilson line, and make a bound state, as we proposed earlier.  The bound state appears at low $z$ in a region where an ff is not intended to be accurate as a description of a real process.  In contrast, the top part of \fref{bound}, for the faster hadrons, matches the corresponding part of \fref{epem} for the physical process.  The bound state of the orphan quark has no need to correspond to any simple observable.

\begin{figure}
  \includegraphics[width=0.4\textwidth]{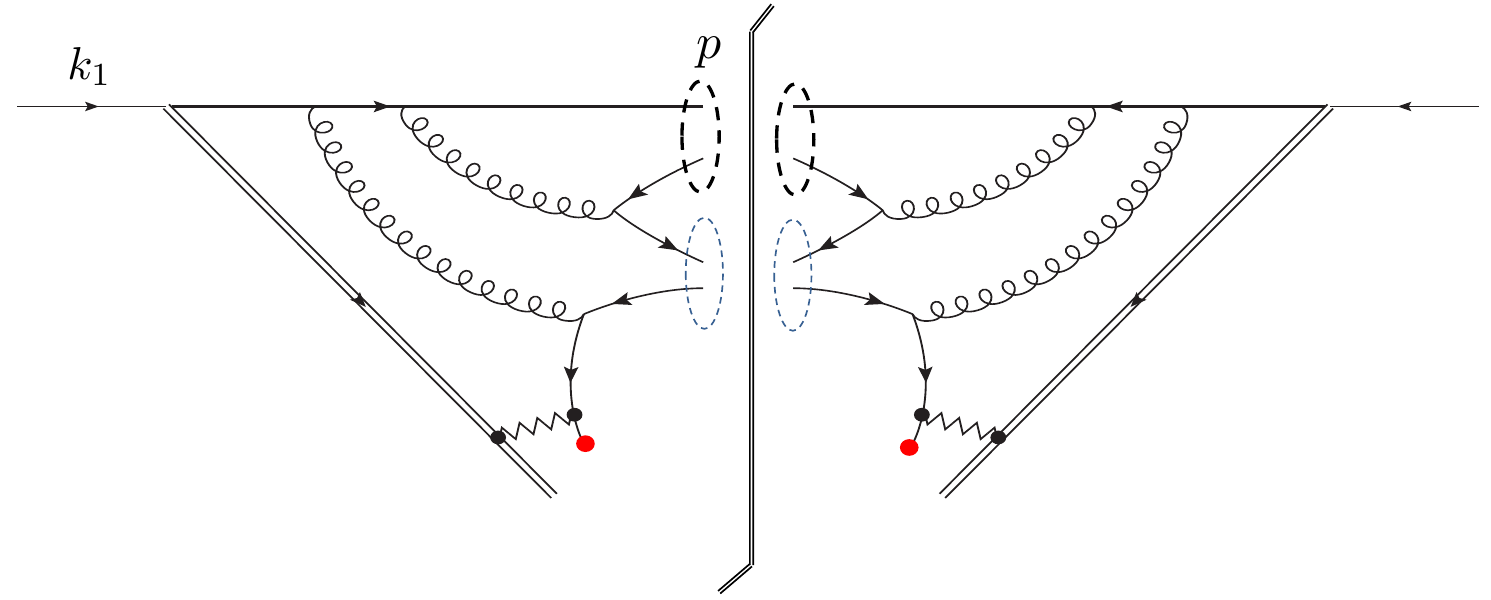} 
\caption{A graph contributing to the calculation of a quark-to-quark ff. The observed hadron is in the thick-dashed black oval. 
The zig-zag line at the bottom is an orphan quark binding to the Wilson line.}
\label{f.bound}
\end{figure}

The simplest way to match ffs to the full process is to implement the afore-mentioned split at central rapidity by replacing the orphan quark by a bound state to a Wilson line that has zero rapidity in the overall center-of-mass frame. In \fref{epem}, this would correspond to an insertion of a time-like Wilson line, with rapidity zero, between the central orphan quark and antiquark.  That would match the definition of a transverse-momentum-dependent ff by Collins and Soper -- see Eq.~(5.1) of Ref.~\cite{Collins:1981uw} -- with its use of an axial gauge and dependence on an auxiliary vector $n$. In contrast, for an integrated ff, the Wilson line is of \emph{infinite} rapidity (opposite to that of the jet).  Given the physical picture just discussed, that would appear at first sight to include too much of the final state.  However, this problem is resolved by a careful analysis of the definitions in the context of a factorization proof, as in \cite{Collins:2011qcdbook}.  First, transverse-momentum dependent ffs are defined, and include ``soft factors'' defined in precisely such a way as to compensate incorrect treatment and overcounting of emission at low rapidity. When one makes the transition to integrated ffs, the soft factors give unity, because of an exact cancellation between real and virtual emission. This type of cancellation between graphs with different kinds of final states highlights the importance of the level of inclusivity in determining the validity of a particular factorization formula.

\section{Implications for sum rules} 
Determining the orphan quark ffs requires information from nonperturbative QCD, and a careful treatment is beyond the scope of this paper. However, general physical considerations lead to some reasonable conjectures that may be useful points of departure for more sophisticated treatments in the future.  

The Wilson line that needs to be included with each field in \eref{basicdef2} is in the light-like plus direction, opposite the nearly light-like minus momentum of the hadronizing quark. It carries the memory of the oppositely moving antiquark (and its associated orphan partons) that appeared in the full process before factorization. It has color opposite to the original quark and ensures that the entire system is color neutral.
Therefore, the Wilson line forms a type of bound state with an orphan quark, as discussed in the previous section. (See \fref{bound}.)  The orphan quark will always have a $z$ momentum fraction almost equal to zero, so the deficit ffs should be approximated by
\begin{equation}
\label{e.conjec1}
d_{{b}/j}(z) \approx c_{j,b} \delta(z) \, , 
\end{equation}
with constant coefficients $c_{j,b}$ that can be different for each type of $b$. In cases with baryon production like \fref{baryon}, it would be a pair of orphan antiquarks that bind with the Wilson line. In this picture, the Wilson line acts as an external source. 
Because the ffs need to satisfy a DGLAP equation,
a literal $\delta$-function may not be appropriate here, but for now \eref{conjec1} should be interpreted simply as saying that an orphan ff is only non-negligible in a very narrow region around $z \approx 0$ and is normalized to a constant $c_{j,b}$. This also matches the physical intuition provided by the discussion of \fref{epem}. In this picture, the last term vanishes for the momentum sum rule in \eref{the_sum_rule2}, giving back the original momentum sum rule of \cite{Collins:1981uw}. But the number sum rules need corrections.

To interpret the meaning of the deficit ffs, it is also useful to consider the consequences of the approximation where baryons contribute a negligible amount to the total number of final states. Then the hadronic sector of the final state always has quark number exactly zero, and the orphan contribution is always just a single quark of flavor $j$ as in \fref{ffs_orphan}. We express this below by writing $\{ b\} \approx \{q_j\}$ where $q$ represents just a single isolated orphan quark of flavor $j$, bound to the Wilson line.
In such a scenario, the deficit ffs satisfy a simple number sum rule,
\begin{equation}
\label{e.quark_num}
\sum_{j'}  \int_0^1 \diff{z} d_{{q_{j'}}/j}(z) = 1 \, ,
\end{equation}
where now $j'$ runs over the flavors of a single orphan quark.
Then the modified hadron number sum rule in \eref{hadron_sum_rule} also takes the very simple form,
\begin{equation}
\label{e.hadron_sum_rule2}
\sum_{h \in \left\{H \right\}} \int_0^1 \diff{z} \, d_{h/j}(z) = \langle N \rangle - 1 \, .
\end{equation}
To satisfy \eref{quark_num}, the coefficients in \eref{conjec1} must obey
\begin{equation}
\label{e.c_sum}
\sum_{j'} c_{j,j'} = 1 \, .
\end{equation}

The corrected expression for the hadron number sum rule in \eref{hadron_sum_rule2} highlights an ambiguity in the interpretation of the total number of hadrons associated with an ff as in \fref{epem}: It is ambiguous whether the central, boxed hadron should be grouped with the left (quark) or right (antiquark) ff. If we insist that it goes with the left side ff, then the number of total particles counted in the quark ff includes the hadron with the orphan quark, and $\langle N \rangle$ would be the appropriate quantity to associate with the total particle multiplicity in the quark ff. However, if the central hadron goes with the right side antiquark ff, then the left ff only includes three hadrons, and $\langle N \rangle - 1$ would be the appropriate measure of multiplicity in the quark ff. 
If $\langle N \rangle \gg 1$, then \eref{hadron_sum_rule2} recovers the hadron number sum rule in, for example, \cite{Majumder:2004br} because the effect of adding or removing just one particle becomes negligible. This picture becomes more complicated once we allow for baryons. However, the no-baryon approximation illustrates how the intuitive hadron number sum rule interpretation might be recovered at high enough energies that the average hadron multiplicity is very large.   

To extend this model still further, and still using the meson dominance model above, we note that it is plausible that 
the $c_{j,j'}$ coefficients in \eref{conjec1} are approximately equal for all active partons $n_f$, given that the orphan 
quark will be very separated in rapidity from the initial quark. In that case, \eref{c_sum} gives
\begin{equation}
c_{j,j'} \approx 1/n_f
\end{equation}
and the charge sum rule in \eref{charge_sum_rule} becomes simply
\begin{equation}
\label{e.charge_sum_rule2}
\sum_{h \in \left\{H \right\}} \mathcal{Q}_h \int_0^1 \diff{z} \, d_{h/j}(z) \approx \mathcal{Q}_j - \frac{1}{n_f} \sum_{j' \in \text{active}} \mathcal{Q}_{j'} \, .
\end{equation}
In this approximation, if the only active quark flavors are only $u$,$d$, and $s$ then the sum of charges is zero and the subtracted term in \eref{charge_sum_rule2} vanishes. Then the standard charge sum rule from \cite{Collins:1981uw} is recovered. If the active flavors include the charm quark, then \eref{charge_sum_rule2} becomes
\begin{equation}
\label{e.charge_sum_rule3}
\sum_{h \in \left\{H \right\}} \mathcal{Q}_h \int_0^1 \diff{z} \, d_{h/j}(z) \approx \mathcal{Q}_j - \frac{e}{6} \, .
\end{equation}

\section{Comments}
The simple model above helps to clarify the meaning of the orphan quark ffs, and it shows how the standard ff sum rules might be approximately true despite the mismatch of final state quark numbers.  But it is important to keep in mind that a deeper understanding of the final states and the nonperturbative properties of the deficit ffs is necessary before this picture can be placed on a very firm footing.  
As it currently stands, it is possible that the range of $z$ where the orphan ffs are non-negligible extends to somewhat higher values than might be expected on the basis of the intuition sketched above, such that there are non-negligible violations of the momentum sum rule. That possibility is especially relevant at moderate hard scales where the kinematical range of validity of factorization is more limited than at very high scales, and the number of final state hadrons is smaller.  

The ff momentum sum rule is rarely ever used \emph{directly} to constrain ffs phenomenologically since to do so requires knowledge of ffs for all hadron flavors, and these are not known with enough precision and over a wide enough range of $z$ for the momentum sum rule to be practically useful~\cite{Kniehl:2000hk,deFlorian:2007aj,Albino:2008fy,Bertone:2017tyb,Abdolmaleki:2021yjf}. To the extent that it is used, it is typically only in the form of an upper bound, 
\begin{equation}
\label{e.bound}
\sum_{h \in H} \int_0^1 \diff{z} \, z \, d_{h/j}(z) \leq 1 
\end{equation}
used to test general consistency, usually with a lower bound on the $z$ integration. Our analysis indicates that this bound remains valid. Nevertheless, given that the sum rule is widely quoted as a fundamental property of ffs (see, for example, Eq.~(19.3) of \cite{ParticleDataGroup:2022pth}), it is important to recognize that, from a theory standpoint, it is not a guaranteed identity. 
Unless a model like \eref{conjec1} is at least approximately valid, then the true upper bound in \eref{bound} might be more or less restrictive. 
It is possibly relevant for existing phenomenology that DGLAP evolution only preserves the momentum sum rule if all the relevant final states appear in the sum, including the orphan quark ffs. An incomplete sum over final state particles is not guaranteed to be preserved under evolution. Practical difficulties with implementing momentum sum rules in combination with DGLAP evolution have been noted in the literature~\cite{Kretzer:2000yf}.  

To our knowledge, the charge sum rule has not been used in applications to phenomenological extractions. 

With regard to the hadron number sum rule in \eref{hadron_sum_rule2}, it may seem that subtracting 1 is a minor modification. However, in semi-inclusive deep inelastic scattering measurements at moderate $Q$ at facilities like Jefferson Lab, typical hadron multiplicities are around $5$, and at a future EIC are expected to be about $12$ to $13$~\cite{JLabprivate}. Adding or removing a single hadron could significantly impact a hadron number interpretation of ffs in scenarios like these. 

What is perhaps more relevant than the impact on existing phenomenology is the role of sum rules in guiding the formulation of new types of 
ffs and establish their interpretation. For example, specific definitions for dihadron ffs and even $n$-hadron ffs were proposed in \cite{Pitonyak:2023gjx} according to a requirement that they satisfy extended versions of the hadron number sum rule. There, the problems discussed in this paper are exacerbated because the $n$-hadron ffs occupy a larger part of the final state phase space, and therefore adding or removing a single particle has a larger impact. Likewise, ~\cite{Accardi:2019luo,Accardi:2020iqn} suggests relating ff sum rules to the dynamical generation of quark masses and jet functions.
In~\cite{Schafer:1999kn,Meissner:2010cc}, momentum sum rules are derived that involve first taking transverse moments of transverse momentum dependent ffs.
For these types of sum rules, the orphan quark problem is compounded by ultraviolet divergent transverse momentum integrals. Finally, the sum rules for fracture functions~\cite{Trentadue:1993ka,Anselmino:2011ss} suffer from the same complication as ffs. 
In all the above applications, a careful look at the issues discussed in this paper is warranted.

Models used in Monte Carlo event generators (e.g., Refs.~\cite{Andersson:1998tv,Kerbizi:2018qpp,Kerbizi:2020uao}) could potentially provide frameworks for clarifying what is needed. There, one is forced to deal directly with descriptions of complete final states. For example, in Ref.~\cite{Ito:2009zc}, the authors find that their model only preserves the momentum sum rule exactly in the limit of infinite final state cascades.

Finally, we propose that understanding the nonperturbative features of the orphan quark ffs through their operator definitions will help to clarify the connection between ffs and full descriptions of the final state hadronization process, and that this may help with the development of applications like those listed above. 
\vspace{.1in}

\acknowledgments
We thank J.~O.~Gonzalez-Hernandez for very useful comments on the text. 
The ideas in this paper were originally inspired by a reading of the recent work in \cite{Pitonyak:2023gjx}, and we thank the authors for a discussion of their paper. T.~Rogers also thanks the Jefferson Lab QCD study group for helpful discussions of relevant topics. 
This work was supported by the U.S. Department of Energy, Office of Science, Office of Nuclear Physics, under Award Number DE-SC0018106. It 
was also supported by the DOE Contract No.\ 
DE-AC05-06OR23177, under which 
Jefferson Science Associates, LLC operates Jefferson Lab. 

\bibliography{bibliography}

\appendix

\begin{widetext}

\section{Calculations in non-gauge model}
To make the sum rules in \erefs{the_sum_rule2}{hadron_sum_rule} less abstract, it is instructive to 
validate them in a renormalizable nongauge theory. We will use a real scalar Yukawa theory with one flavor of Dirac fermion ``quark'' and one pion. Both fields will have nonzero masses, but we will set the masses equal to simplify calculations, $m_\text{quark} = m_\pi = m$. We will consider the case of a single quark flavor $j$. Normal Feynman graph calculations give the pion-in-quark ff from the definition in \eref{final_state_def} in $\msbar$ renormalization at lowest nontrivial order,
\allowdisplaybreaks 
\begin{align}
d_{\pi/j}(z;\mu) \;
&{}= \picineq{FFgreal}  + \text{\msbar \; C.T.} \no
&= a_\lambda(\mu) z \left[ \ln \parz{\frac{1}{1-z + z^2}} + \ln \frac{\mu^2}{m^2} - 1 + \frac{(2-z)^2}{1-z + z^2 } \right] \, , \label{e.realNLOpion}
\end{align}
where $a_\lambda(\mu)$ is a coupling constant and $\mu$ is the usual $\msbar$ scale from dimensional regularization. The term with a final-state quark is, to the same order,
\begin{align}
&d_{q_j/j}(z;\mu) \no
&{} = \picineq{FFlo} + \left[ \picineq{FFbubble}  + \text{H.C.} \right] + \picineq{FFqreal} + \text{\msbar \; C.T.'s}  \no 
&{}= \delta(1 - z) - a_\lambda(\mu) \delta(1 - z) \left[\pi \frac{\sqrt{3}}{2} -2  + \ln \frac{\mu}{m} \right] + a_\lambda(\mu) (1-z) \left[ \ln \parz{\frac{1}{(1-z)^2 + z}}  + \ln \frac{\mu^2}{m^2} - 1 + \frac{(1+z)^2}{(1-z)^2 + z} \right] \, .
\label{e.ff.to.quark}
\end{align}
Taking the zeroth and first moments gives
\begin{align}
&\int_0^1 \diff{z} d_{q_j/j}(z;\mu) = 1 - a_\lambda(\mu)  \left[\pi \frac{\sqrt{3}}{2} -2  + \ln \frac{\mu}{m} \right] + a_\lambda(\mu)  \left[\pi \frac{\sqrt{3}}{2} -2  + \ln \frac{\mu}{m} \right] = 1 \, , \label{e.qzeromom} \\
&\int_0^1 \diff{z} d_{\pi/j}(z;\mu) = a_\lambda(\mu)  \left[\pi \frac{\sqrt{3}}{2} -2  + \ln \frac{\mu}{m} \right]  \label{e.pizeromom} \, , \\
&\int_0^1 \diff{z} z d_{q_j/j}(z;\mu) = 1 - a_\lambda(\mu)  \left[-\frac{13}{9} + \frac{\pi }{\sqrt{3}} + \frac{1}{3} \ln \frac{\mu^2}{m^2} \right] \label{e.qonemom} \, , \\
&\int_0^1 \diff{z} z d_{\pi/j}(z;\mu) = a_\lambda(\mu)  \left[-\frac{13}{9} + \frac{\pi }{\sqrt{3}} + \frac{1}{3} \ln \frac{\mu^2}{m^2} \right] \, . \label{e.gonemom}
\end{align}
Equation \eqref{e.qzeromom} demonstrates that the quark number sum rule in \eref{quark_num} is indeed satisfied. Noting that the pion has zero charge, \eref{qzeromom} also confirms the charge sum rule in \eref{charge_sum_rule}. 
Finally, substituting \eref{qonemom} and \eref{gonemom} into \eref{the_sum_rule2} shows that
the momentum sum rule in \eref{the_sum_rule2} is satisfied, and it appears as 
\begin{equation}
\label{e.the_sum_rule2_model}
\left\{ a_\lambda(\mu)  \left[-\frac{13}{9} + \frac{\pi }{\sqrt{3}} + \frac{1}{3} \ln \frac{\mu^2}{m^2} \right] \right\} = 1 - \left\{1 - a_\lambda(\mu)  \left[-\frac{13}{9} + \frac{\pi }{\sqrt{3}} + \frac{1}{3} \ln \frac{\mu^2}{m^2} \right] \right\} \, .
\end{equation}
To match the formula in QCD, with its deficit term that involves final-state particles with quark quantum numbers, we have put the quark term on the right-hand side.

The sum rules are only satisfied here if the contribution of an on-shell final-state quark to the ff is included. Note that if this term were dropped from the right-hand side 
of \eref{the_sum_rule2_model}, not only would the momentum sum rule be violated, but the amount of the violation would depend on the evolution scale $\mu$. An important part of this contribution is from large $z$, as in the delta-function terms in \eref{ff.to.quark}.
Therefore, in this theory, the quark term is an important contribution to the measurable hadronization of a quark-induced jet.  In terms of our notation for QCD, quark particles would be included in the set we denoted by $H$, and there would be no orphan particles to put in the set $B$.

The hadron number sum rule in \eref{hadron_sum_rule2} becomes, using \eref{pizeromom}, 
\begin{equation}
\label{e.numbersum_example}
a_\lambda(\mu)  \left[\pi \frac{\sqrt{3}}{2} -2  + \ln \frac{\mu}{m} \right] = \langle N \rangle - 1 \, .
\end{equation}
If the coupling is negligible, the left side of \eref{numbersum_example} vanishes and the sum rule is 
just $\langle N \rangle = 1$, i.e. the final state is always just a single quark. 

\end{widetext}

\end{document}